\documentclass[a4paper]{article}
\usepackage{graphicx}

\begin{document}

\begin{center}
{\bf \Large
Cops or robbers - a bistable society}
\bigskip

{\large
K. Ku{\l}akowski$^{\dag}$
}
\bigskip

{\em
Faculty of Physics and Applied Computer Science,
AGH University of Science and Technology,
al. Mickiewicza 30, PL-30059 Krak\'ow, Poland\\

}

\bigskip

$^\dag${\tt kulakowski@novell.ftj.agh.edu.pl}

\bigskip
\today
\end{center}

\begin{abstract}
The norm game described by Axelrod in 1985 was recently treated with the master equation formalism. 
Here we discuss the equations, where {\it i)} those who break the norm cannot punish and those who
punish cannot break the norm, {\it ii)} the 
tendency to punish is suppressed if the majority breaks the norm. The second mechanism is new. 
For some values of the parameters the solution shows the saddle-point bifurcation.
Then, two stable solutions are possible, where the majority breaks the norm or the majority punishes.
This means, that the norm breaking can be discontinuous, when measured in the social scale.
The bistable character is reproduced also with new computer simulations on the Erd{\H o}s--R\'enyi directed 
network.  
\end{abstract}



\bigskip

\section{Introduction}

In various applications of game theory, the problem of the timescale is crucial. In biology, genetic changes appear to be visible during geological times. The utility function is the survival fitness and the agent is a species. In economy, the relaxation time can be as short as a change of prices on the stock exchange. The utility is money and the agent is a broker. In social sciences, one can consider historical processes, as rise and fall of empires; this can take centuries. Then the utility can be the maintenance of the political existence of a nation, and the agent is represented by a series of consecutive ministers of foreign affairs. However, people older than thirty can easily indicate numerous examples of social changes, which happened to occur in the last ten years. In such cases, the process itself takes place in parallel with variations of the public opinion about it; individual beliefs change. In particular, the payoffs accessible with given strategies change as well. This means that the whole process cannot be fully described within one game. The time evolution of a social norm seems to belong to this class \cite{axe1}. Paradoxically then \cite{stan}, game theory appears to be applicable in very long timescale or in short timescale, when the agent's wisdom is encoded in his genes or in his palmtop. The case of the human brain remains the challenge. As it was thoroughly discussed by Robert Axelrod, here the concept of evolutionary thinking provides appropriate frames. The situation can be even more complex if we take into account that some social norms are actually contrary to an individual's interest. Such a norm either remains constant or evolves so slowly that one generation transmits it to the other in seemingly the same form. The seminal paper of Axelrod \cite{axe0} gives an example: the norm of dueling. Here we are interested in the case when a slow evolution of the norm -- an increase of the cost of punishing -- leads to a decrease of the number of those who obey the norm. We imagine that an i
ncrease of the number of defectors, i.e. those who break the norm, leads to a decrease of the number of punishers, what in turn makes is easier to defect -- this is a positive feedback which is necessary if two different phases are to be observed.

The goal of this work is to demonstrate the existence of this bistability within two simple methods. The first one is the mean field master equation, and the second -- the stochastic dynamics of probabilities of strategies of agents placed at nodes of a random network. In both cases, there are two conditions which seem to be necessary to get the bistability. The first is that it should be impossible for the same agent to break the norm and to punish. In the social reality, there are some norms where indeed this condition is true: it should be impossible to revile drinking if one himself abuses alcohol. On the other hand, a traffic policeman can drive after a beer or three, not to look for other examples. Then our considerations will apply only to the norms where the above condition is preserved. The second is that the attitude to punish decreases with the number of defectors. This also is not always the case, but it seems natural that once the majority defects, the punishing becomes out of fashion. These two conditions are built into the methods used here.

We note that several authors have continued the Axelrod approach and some simulations can be found in the literature \cite{hau,perr,sob,gaizq}. Often the authors refer to the idea of a metagame; a norm is to be supported not only by punishing for its defection, but also by punishing for not punishing. This
idea was considered by Axelrod to be a driving force to an ultimate existence of norms. In this text, the metagame is absent; still here and previously
\cite{moje} we demonstrate that norms can be preserved in the social scale without this additional mechanism. Up to the knowledge of this author, yet the social bistability has not been systematically investigated.

\begin{figure}
\includegraphics[scale=0.50,angle=-90]{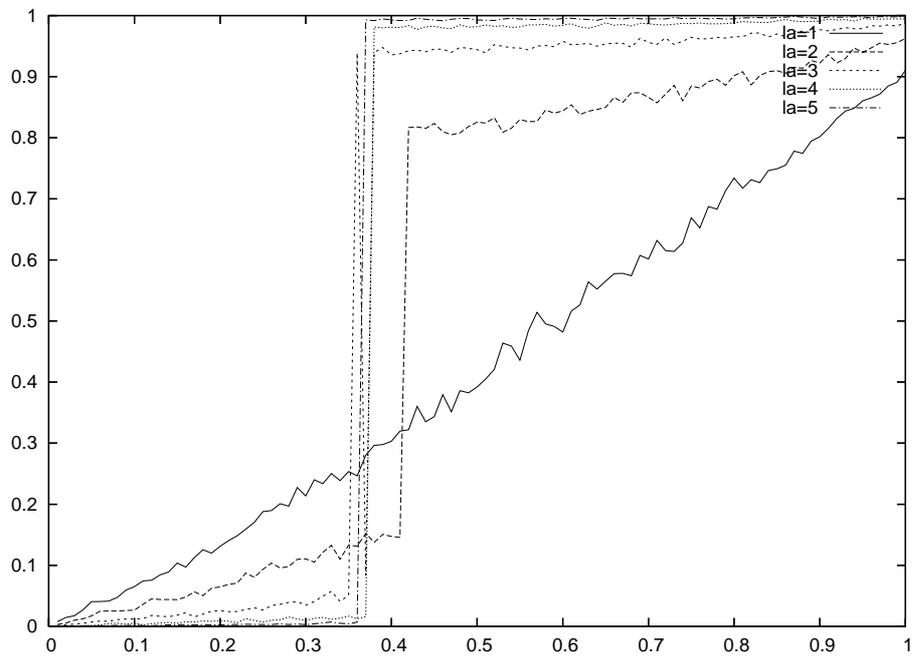}
\caption{The stationary probability $z$ of defection as dependent on the initial boldness $\rho$ for various values of the average degree $\lambda$ of the network nodes. For $\lambda=2$, the number of timesteps is $2\times 10^4$, for higher $\lambda$, $10^3$ is long enough. For $\lambda=1$, the obtained 
plot slightly fluctuates with the number of timesteps.}
\end{figure}

\begin{figure}
\includegraphics[scale=0.50,angle=-90]{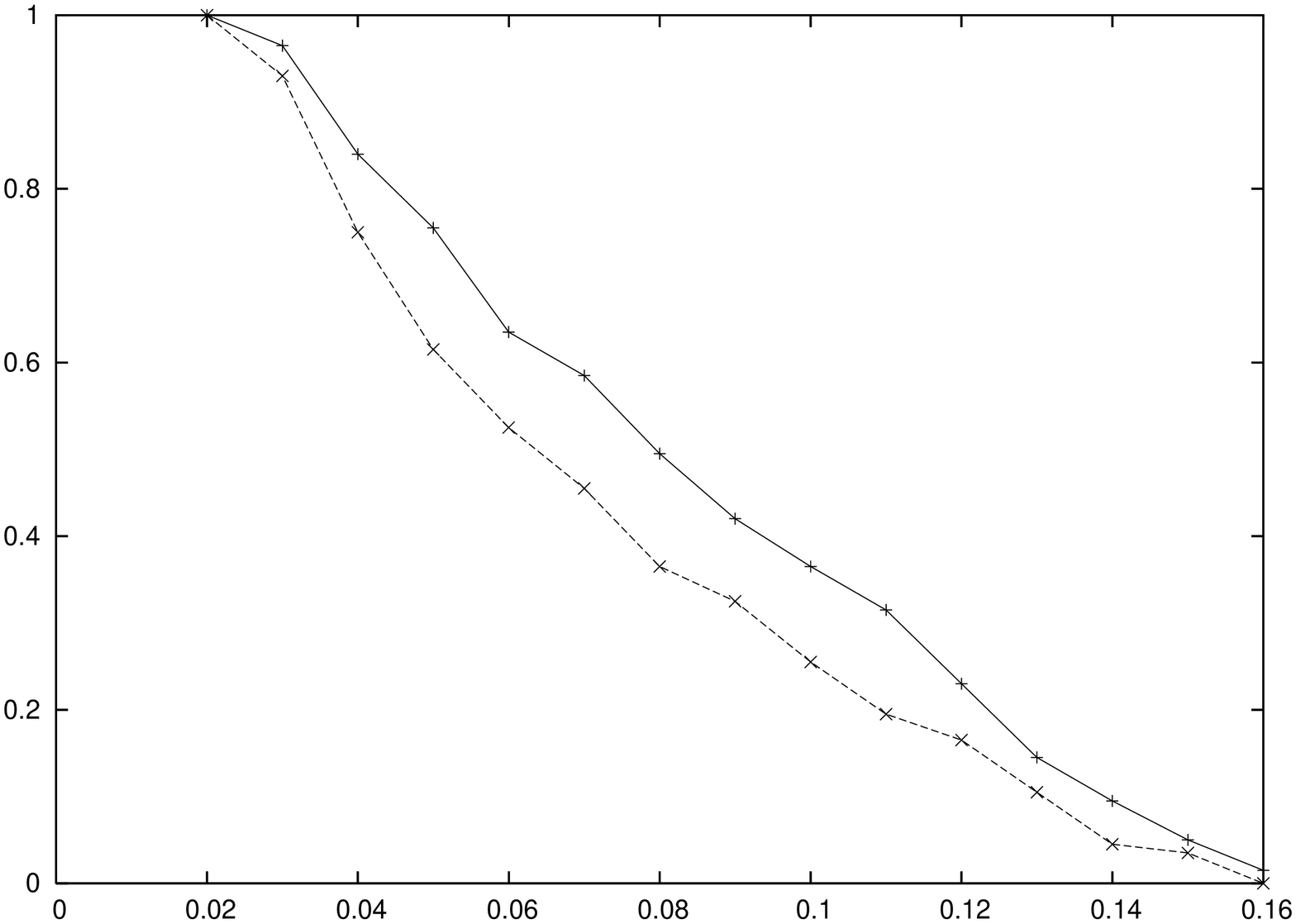}
\caption{The boundary $\rho_c(\gamma)$ between the bassins of attractions of states 'most punish' (below the curve) and 'most rob' (above the curve). In these states, $<z>\approx 0$ and $1$, respectively. As a rule, $<x>\approx 0$ in the stationary state. The plots are obtained for $10^4$ timesteps, $N=4\times 10^4$, $\lambda =5$, $\beta=0.1$. For the upper curve $\mu=1.0$, and for the lower one $\mu=2.0$.}
\end{figure}

In the next section the mean field master equations are explained. Although the mean field theory neglects correlations between agents, its simplicity and transparentness makes it a necessary prerequisite of any simulation of collective phenomena. Our equations provide maybe the simplest method to cope with the problem. The parameters used refer to the evolution of the probabilities of norm breaking and of punishing due to the previous history of the agents. In this way we omit the parameters of payoffs, which turn out to be unnecessary. We show that in some range of the parameters, the stable fixed points of the master equations display the bistability. In the third section we describe the simulation scheme. The punishment can be executed between the agents placed in the Erd{\H o}s--R\'enyi network, along the directed links. Then, the number of links incoming to a node is equivalent to a number of observers of an agent at this node; it is only these observers who can punish. Here more details are arbitrary and the study is less complete. Still we show that the bistability exists if the average number of nearest neighbours (the node degree) is large enough. The last section is devoted to discussion.

\section{The master equations}

The equations below are just a slight modification of the model (B1) in Ref. \cite{moje}. This model contains an assumption that it is impossible to  break the norm and punish simultaneously. This is the first of the two conditions listed in the Introduction. The equations are

\begin{equation}
\frac{dz}{dt}=ax-bzy 
\end{equation}
\begin{equation}
\frac{dy}{dt}=-cyz+exz(1-z)
\end{equation}
where $y$ and $z$ are the probabilities of defection and punishing, respectively; $x=1-y-z$ is the probability of just obeying the norm and not punishing.
In terms of Axelrod, $z$ and $y$ refer to the boldness and the vengeance, respectively. In Eq. 1, $a$ is the rate of increase of $z$ per agent because of the gain which he gets when he breaks 
the norm, and $b$ is the opposite rate because of the inhibiting results of the punishment. We can term these rates as 'temptation' and 'punishment'.
The second term on the r.h.s. term is proportional to $y$, because a punisher is necessary here. In Eq. 2, the first term on the r.h.s. describes
a decrease of the attitude to punish; this is the 'cost' term. The last term could be termed 'rage' as it reflects an increase of the tendency to punish; obviously it is proportional
to the number of non-punishers $x$, because it is only them who can start to punish, and to the number of defectors. The second modification noted above is that the last term is also proportional to $1-z$. The introduction of this modification reflects an exhaustion felt by a punisher when he sees that so many people defect. By this interpretation, all constants $a$, $b$, $c$ and $e$ are positive. Without loss of generality we can put $e=1$; this just fixes the timescale and has no influence on the properties of the fixed points.

The fixed point $(x^*,y^*,z^*)=(0,1,0)$ is always unstable, and the one $(0,0,1)$ is always stable; in the latter case the Jacobian has two 
eigenvalues $-a,-c$. There are also other solutions, existing only if $4ac<b$:

\begin{equation}
z^*=\frac{1}{2}\Big(1\pm\sqrt{1-\frac{4c}{\phi}}\Big)
\end{equation}
\begin{equation}
y^*=\frac{1-z^*}{1+\phi z^*}
\end{equation}
where $\phi=b/a$. The fixed point with $"+"$ is always unstable and the other with $"-"$ is always stable. At the point $\phi=4c$ both $"+"$ and $"-"$ merge and disappear; then we have the saddle-point bifurcation \cite{glen}.

This bifurcation seems to be generic at least in the sense discussed in Ref. \cite{moje} (model B2), where the cost term was $-cy$ instead of $-cyz$. Then the bifurcation point is shifted from $\phi=4c$ to $\phi=27c/4$. The fixed point $(0,0,1)$ is again always stable. This behavior is contrary to the models
(A1) and (A2) discussed in Ref. \cite{moje}, where  the introduction of the exhaustion factor $1-z$ to the rage term leads to the transcritical bifurcation. The difference between the models (B1,B2) and (A1,A2) is that in the latter the same agent can defect and punish. We 
deduce that the discontinuous jump of the boldness $z$ at the bifurcation point does appear only if the above "schizofrenic" behavior is excluded.

\section{The simulation}

At first we construct the random Erd{\H o}s--R\'enyi directed network of $N$ nodes. For each node $i$, an integer number $nn(i)$ is selected randomly
from the Poisson distribution with the mean value $\lambda$, the same for the whole network. Next, $nn(i)$ nodes are selected randomly and linked 
to $i$-th node; the punishment can be executed only via the links, preserving their initial direction. To each node two variables are assigned, i.e. the probability of norm breaking (boldness) $z(i)$ and the probability of punishing $y(i)$ (vengeance). Initial values of $z(i)$ are selected randomly
from the range $(0.9\rho,\rho)$. Roughly, $\rho$ can be treated as a measure of the initial boldness. Similarly, initial values of $v(i)$ are selected 
as $(1-z(i))/\mu$, with $\mu >1$.

During the simulation, an agent is updated at a randomly selected node $i$. The agent either obeys the norm or defects with probabilities $1-z(i)$ or 
$z(i)$, respectively. In the latter case its boldness $z(i)$ is set to 1 and its vengeance $y(i)$ is set to zero. Then its neighbors are updated one by one; they either abstain from punishing the defector or punish with probabilities $1-v(i)$ or $v(i)$, respectively. Once a neighbor $k$ abstains from punishing, its characteristics becomes the same as the one of the defector: $z(k)$ is set to 1 and $y(k)$ to zero. 

Once a neighbour $k=j$ punishes the defector $i$, the boldness of $i$ is reduced according to the rule

\begin{equation}
z(i) \to (1-\beta)z(i)
\end{equation}
where $\beta$ is the punishment factor from the range $(0,1)$. Simultaneously the vengeance of the punisher $j$ decreases according to the rule

\begin{equation}
y(j) \to (1-\gamma)y(j)
\end{equation}
where $\gamma$ is the cost factor, also from the range $(0,1)$. Then, the system dynamics is determined by the parameters $\beta$ and $\gamma$.

The simulation was performed for $N$ from $10^3$ to $5\times10^3$, $\lambda$ from 1 to 5. One step of the simulation is equivalent to $N$ updates of randomly selected nodes; in most cases, the stationary state is obtained after $10^3$ time steps. The numerical results indicate, that for certain values of the parameters $\beta,\gamma$ the stationary state depends on the initial state, encoded here as $(\rho,\mu)$. If $\lambda=2$ or more, a sharp change of the stationary value of $<z>$, averaged over the nodes, is observed against $\rho$. This bistable character of $<z>$ is shown in Fig. 1. As a rule,
$<y>$ is close to $1-<z>$; either almost all defect, or almost all punish. There are two exceptions: the first is the case for $\lambda=1$, where the plot is monotonic. The second is the case for $\lambda=2$; there, the maximal value at the jump is reduced by a factor close to $\exp(-2)$, i.e. the number
of nodes without incoming links. As a rule, the boldness and the vengeance of these nodes is either zero or one, it is only their number what changes
continuously with $\rho$ for $\lambda =1$.

In the second plot we show the boundary value $\rho_c$ between two fixed points: $(x^*,y^*,z^*)=(0,1,0)$ and $(0,0,1)$. If our initial state is 
prepared with  $\rho < \rho_c$, the obtained stationary state is equal to $(0,1,0)$ within numerical accuracy. Above $\rho_c$, the system tends to
$(0,0,1)$. As we see, the obtained curves (for different initial values of $y$) differ only slightly. Both of them evolve from unity to zero with
the punishment cost $\gamma$. This means, that for small $\gamma$, the phase 'all punish' is stable, but this stability is gradually lost with 
the increasing cost of punishing. Above some value of $\gamma$, the punishment strategy is not stable. Below some another value of $\gamma$, the defection strategy does not hold. This behaviour is qualitatively similar to the result obtained within the mean field theory, as described in the previous section. Within both of methods, the bistable behavior is observed in some range of the punishment cost.

\section{Discussion}

The point of departure of our models is that the social acceptance of norms could be an example of self-supporting phenomena.
 For a given norm, there is a positive feedback between its individual level of internalization, averaged over the society 
members, and its acceptance at the social scale. In physics, such a positive feedback leads often to a kind of phase 
transition to the phase, where more than one stationary state are possible. The famous Ising model is known to 
be a good tool to simulate such transitions. It is only recently that this model has been used to analyse the tax evasion dynamics 
\cite{zak}. In our model, there are two final states: to break the norm or to punish. This duality is the basic assumption
in the Ising model; here however it is the result and not an assumption, because the third state 'to obey the norm and do not
punish' is not occupied in the stationary phase. Our analytical and numerical results suggest that there are two conditions which seem sufficient 
for the appearance of the bistable phase. The first is the strict separation of the role of norm defector from the role of 
punisher. The second is a decrease of the attitude to punish in the situation when a large part of the society defects. 
Even if both conditions seem obvious, in our opinion they can be subject of sociological research when 
a given norm is considered.

The goal of this paper is that if both conditions are fulfilled, the social acceptance of the norm can display a discontinuous behaviour and a kind
of hysteresis effect. Once the cost term varies slowly, the proportion of those who defect can vary slowly as well until the phase loses its stability.
Then, suddenly, the system jumps to another phase which is the only stable one for the new value of the punishment cost. The hysteresis effect is
equivalent to the dependence of the state on its history; in social problems, such a dependence is a rule rather than an exception. Here it indicates,
that both the state when the norm is obeyed and the state when it is defected have some inherent stability, as there is no continuous transition 
from the one to the other. In particular, the state of lawlessness is more stable than we would like it to be. On the other hand, the number of those who 
obey the norm can reduce in a sudden and unexpected way. 

The discontinuous character of the dynamics can be an obstacle in restoring the state where a given norm is preserved. 
It can be profitable to introduce a temporal mixture of states, as a kind of amnesty or state of grace, when those who had 
broken the norm in the past are not only forgiven but also permitted to punish other defectors. On the other hand, rare 
individuals who punish in the defecting society should be highly gratified; it is also 
good to make them famous, if only their safety can be secured. These theoretical suggestions find a confirmation in the 
common practice.

{\bf Acknowledgements} The author is very grateful to Dietrich Stauffer for years of collaboration and criticism and wishes
him good health and good luck.

\end{document}